\documentclass[twocolumn,showpacs,preprintnumbers,amsmath,amssymb,eps]{revtex4}
\usepackage{graphicx}% Include figure files
\usepackage{dcolumn}% Align table columns on decimal point
\usepackage{bm}% bold math

\begin{document}

%\preprint{APS/123-QED}

\title{Distinguishing the origin of the superconducting state from the
pseudogap of high-temperature superconductors\\}

\author{E. V. L de Mello and Raphael B. Kasal}
%\altaffiliation[]{evandro@if.uff.br}
%\author{Raphael B. Kasal, C. A. Passos}
\affiliation{%
Instituto de F\'{\i}sica, Universidade Federal Fluminense, Niter\'oi, RJ 24210-340, Brazil\\}%
%\author{Otton Teixeira da Silveira Filho}
%\affiliation{Instituto de Computac\~ao, Universidade Federal Fluminense, Niter\'oi, RJ 24210-340, Brazil}

\date{\today}% It is always \today, today,
             %  but any date may be explicitly specified

\begin{abstract}

We consider an electronic phase separation process that generates
regions of different charge densities, or local dopings, as the origin
of the inhomogeneous charge density of high $T_c$ superconductors.
We show that it gives rise to a phase boundary potential between
such doping disordered regions or grains. The Bogliubov-deGennes 
self-consistent calculations in this disordered medium yield position 
dependent superconducting gaps which are, for all dopings, smaller than 
those derived from the local density of states with a pseudogap behavior.
Studying these two sets of gaps for different temperatures
and dopings, we are able to reproduce many 
many observed properties of superconducting cuprates.
This scenario is consistent with a resistivity transition
driven by Josephson coupling among the superconducting grains.

\end{abstract}

% PACS, the Physics and Astronomy
% Classification Scheme.
%\keywords{Suggested keywords}%
\maketitle
%\section{Introduction}
The origin of the superconducting gap associated with the superconducting
state and the relation to the pseudogap above the transition
temperature ($T_c$) remains one of the central questions in
high-$T_c$ research. In this letter, we show
that the solution of this important problem is 
connected to an electronic granular structure derived from
an electronic phase separation (EPS) transition.

There are steadily accumulating evidences that the charge distribution 
is microscopically inhomogeneous in the $CuO_2$ planes of
high temperature superconductors 
(HTSC)\cite{Tranquada,Pan,McElroy,Gomes,Pasupathy,Kato,Pushp,Kato2}.
Low temperature Scanning Tunneling Microscopy (STM) has demonstrated
non-uniform energy gaps $\Delta$ that vary on the length scale of
nm\cite{Pan} over the whole surface of these materials. 
These gaps have two types 
of shapes\cite{McElroy}, and some remain well above the 
superconducting critical temperature $T_c(p)$\cite{Gomes,Pasupathy}. 
More recently, the
existence of two types of energy gaps have been observed on
electronic Raman scattering experiments\cite{LeTacon},
STM data\cite{Kato,Pushp,Kato2}, Angle Resolved Photon Emission 
(ARPES)\cite{Shen} and combined STM-ARPES\cite{Mad}. However,
the origin and even the existence of this two-gap picture is
still a matter of debate\cite{Campuzano,Chatterjee}.

A possible reason to this complex behavior is
an EPS transition driven 
by the minimization of the free energy\cite{Mello09}. This 
minimization is due to the formation of anti-ferromagnetic (AF) 
regions of almost zero local doping ($p_i\approx 0$) with lower 
free energies than the regions with large local doping level 
($p_i\approx 2p$), where $p$ is the average hole doping of the 
sample\cite{Mello09}.  As $p$ increases, the Coulomb repulsion 
in the large local doping  regions generates a high energy 
cost to the phase separation process
and the EPS ceases in the overdoped region, in agreement with the
disappearance of the local AF fluctuations\begin{large}\begin{Large}                                                       \end{Large}                                          \end{large}\cite{Tranquada2}.
This cooperative phenomenon gives rise to (grain) boundary 
potentials in the $CuO_2$ planes, single particle bound states 
and intra-grain superconductivity at low temperature. 
These superconducting regions develop Josephson couplings
among them and the resistivity transition $T_c(p)$
occurs when the Josephson
energy $E_J(p)$ is equal to $K_BT_c(p)$.

\begin{figure}[ht]
\begin{center}

     \centerline{\includegraphics[width=7.0cm]{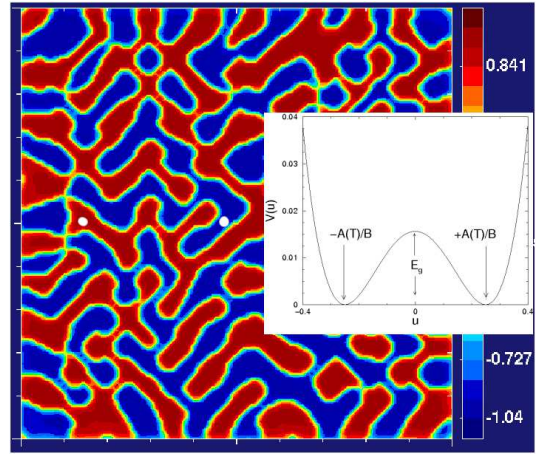}}
%    \vspace{0.45cm}
%    \centerline{\includegraphics[width=5.0cm,angle=-90]{LDOSV7T09.n21Ts.461.ps}}
%   \end{center}
%  \end{minipage}V_{sc
\caption{ (color online) The density map simulation of the 
inhomogeneous charge density on a $100 \times 100$
sites. In the inset, the GL 
potential with the two minima representing the
two (high and low density) solutions\cite{Mello04} and
the potential barrier $V_{gb}$ between them. Marked white
points are located where some calculations shown here were made. }
\label{MapVu} 
\end{center}
\end{figure}

We used the theory of Cahn-Hilliard (CH)\cite{CH,Otton,Mello04}
to binary alloys to describe the EPS transition. The
transition order parameter is the difference between the
local and the average charge density $u(p,i,T)\equiv (p(i,T)-p)/p$.
In Fig.(\ref{MapVu}), we show a typical density map  with
the two (hole-rich and hole-poor) main solutions given by
different colors.

In the inset of Fig.(\ref{MapVu}) we plot the usual potential 
from the Ginzburg-Landau (GL) free energy expansion used 
in the CH equation\cite{Mello09} 
with its two minima and the potential barrier $V_{gb}=V_{gb}(p,T)$ between them.
The barriers generate shallow potential wells in the $CuO_2$ planes that
always have an exponential small bound state\cite{Landau}. 
These bound states lower the
kinetic energy to allow, under favorable conditions, local
hole pair formation in the antiadiabatic limit\cite{JR}.

The calculated  $u(i,T)$ (or $p(i,T)$) density map, as
shown in Fig.(\ref{MapVu}), is used as
the {\it initial input} and it is  maintained fixed throughout the
self-consistent Bogoliubov-deGennes (BdG) calculations. Notice that
the charge inhomogeneity $p(i,T$ and $V_{gb}(p,T)$ are 
correlated. In what 
follows, the parameters involved in $V_{gb}(p,T)$ are chosen to 
match the average local density of states
(LDOS) gaps measured by low temperature STM  on $0.11 \le p \le 0.19$
Bi2212 compounds\cite{McElroy}.  All others parameters are 
similar to values previously used\cite{DDias08,Mello08}. 

The effect of the temperature in the potential is taken
into account using $V_{gb}(T)\sim (1-(T/T_{PS})^{1.5}$, 
as demonstrated by  Cahn and Hilliard\cite{CH}. Thus we 
were able to obtain the intra-grain or {\it
local superconducting temperatures} $T_c(i)$, i.e., the onset
temperature for the d-wave superconducting gap 
$\Delta_d(i,T)$ at a given location $i$. The largest
value of all $T_c(i)$ determines the temperature $T_{on}(p)$ that
marks the onset of local superconductivity of the sample. 

Similar to granular superconductors\cite{Merchant}, the
superconducting transition occurs in two steps: first by the
appearing of intra-grain superconductivity and than by Josephson coupling
with phase locking at a lower temperature. This approach {\it
provides a clear interpretation to the superconducting amplitude
and the  measured quasiparticles dispersion above  
$T_c(p)$}\cite{Campuzano,Chatterjee}.

\begin{figure}[ht]
\begin{center}
  \begin{minipage}[b]{.1\textwidth}
    \begin{center}
%     \centerline{\includegraphics[width=6.0cm,angle=-90]{GnuP.EjxT4.ps}}
     \centerline{\includegraphics[width=7.0cm]{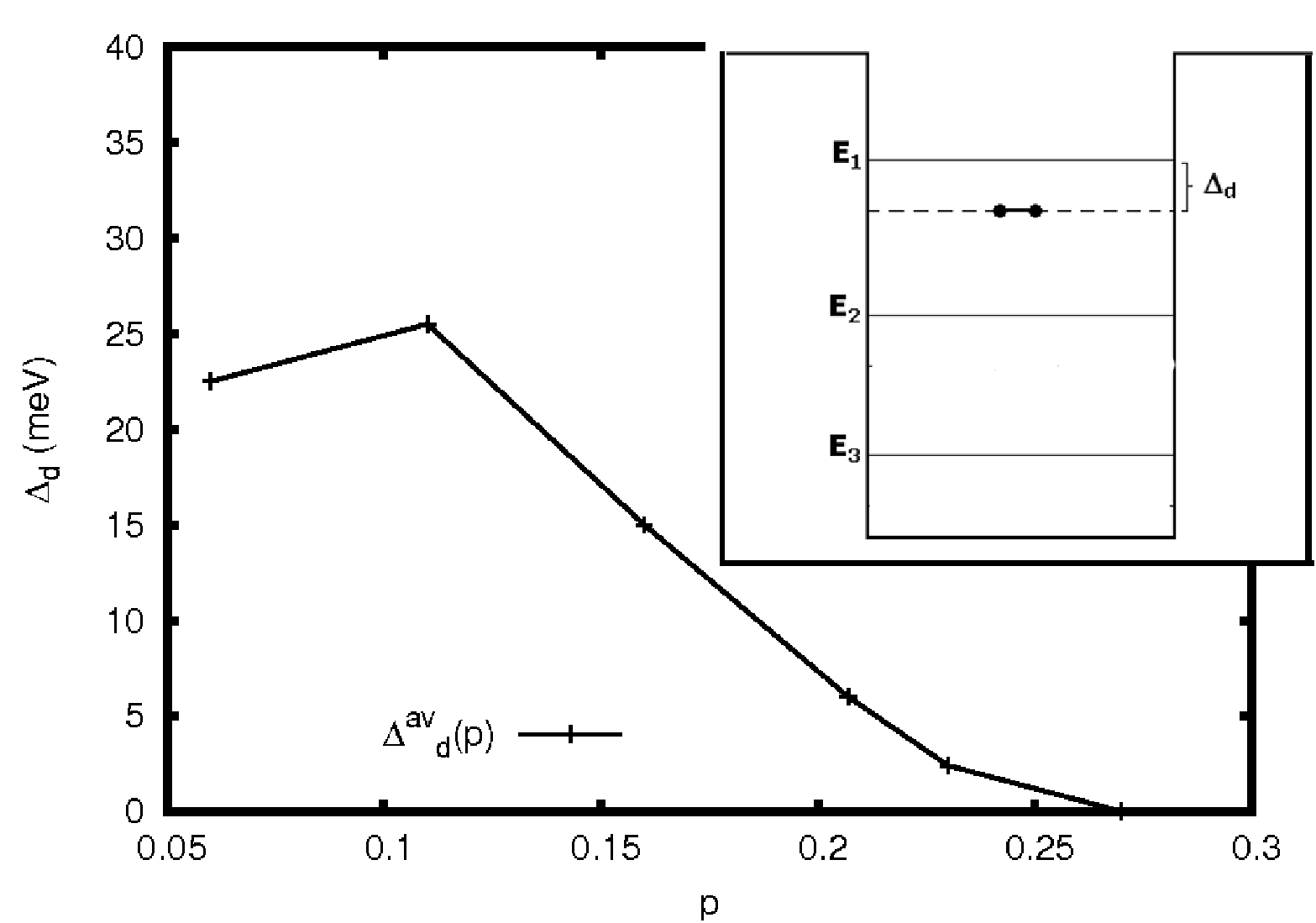}}
%    \vspace{0.45cm}
    \centerline{\includegraphics[width=7.0cm]{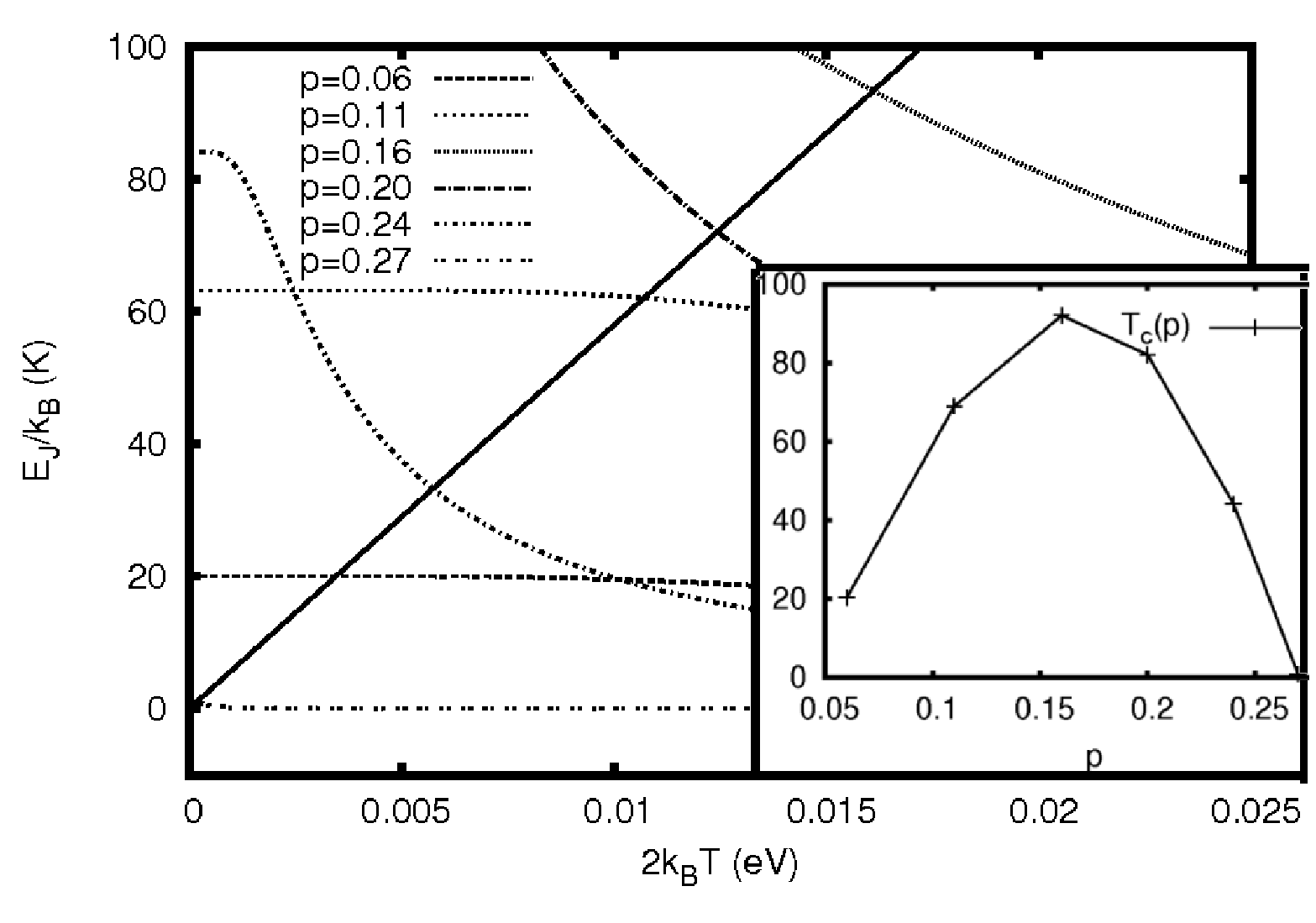}}
    \end{center}
  \end{minipage}
\caption{Top panel, the average $\Delta_d(p)$, and in the
inset, the schematic single particle and superconducting energy levels 
in a grain. In the low panel, the thermal energy
$k_BT$ and the Josephson coupling among superconducting grains $E_J(p,T)$
for some selected doping values as function of T. The intersections
give $T_c(p)$, as plotted in the inset.  }
\label{EJTc}
 \end{center}
\end{figure}

By using the theory of granular superconductors\cite{AB} to these
electronic grains and the calculated average superconducting
amplitudes $\Delta^{av}_d(T,p)\equiv \sum_i^N \Delta_d(T,i,p)/N$, where N is
the total number of sites, we can estimate the values of $T_c(p)$.
\begin{eqnarray} E_J(p,T) = \frac{\pi h}{4 e^2 R_n}
tanh(\frac{\Delta^{av}_d(T,p)}{2K_BT_c}).
\label{EJ} 
\end{eqnarray} 
Where
$\Delta^{av}_d(T,p)$ is the average of the local superconducting gaps
$\Delta_d(i,p,T)$ on a $N \times N$ ($N=28$, $36$ and $42$) square
lattice which are plotted in the top panel of Fig.(\ref{EJTc}). The inset
shows the schematically single particle levels at a shallow 
puddle whose  walls are proportional to $V_{gb}$. $R_n$ is 
the normal resistance of a given compound, which
is proportional to the planar resistivity $\rho_{ab}$ 
measurements\cite{Takagi} on the
$La_{2-p}Sr_pCuO_2$ series. In the low panel of Fig.(\ref{EJTc}),
the Josephson coupling $E_J(p,T)$ is plotted together with the
thermal energy $k_BT$ whose intersection yields the critical temperature
$T_c(p)$, as shown in the inset. The
values are in reasonable agreement with the
Bi2212 $T_c(p)$, as expected, since $V_{gb}$ was chosen to 
match the Bi2212 low temperature LDOS\cite{McElroy}.

In the BdG
approach, the symmetric local density of states (LDOS) is
proportional to the spectral function\cite{Gygi} and may be written
as \begin{eqnarray}
 N_i(T,V_{gb},eV)&=&\sum_n[|u_n({\bf
x}_i)|^2 + |v_n({\bf x}_i)|^2]\times  \nonumber \\
&& [f_n^{'}(eV-E_n)+f_n^{'}(eV+E_n)]. \label{LDOS}
\end{eqnarray}
The prime is the derivative with respect to the argument. 
$u_n, v_n$ and $E_n$ are respectively the eigenvectors and
eigenvalues of the BdG matrix
equation\cite{Mello04,DDias08}, $f_n$ is the
Fermi function and $V$ is the applied voltage. 
$N_i(T,V_{gb},eV)\equiv LDOS(V_{gb})$ is proportional to the
tunneling conductance $dI/dV$, and we probe the effects of the 
inhomogeneous field by examining the ratio $LDOS(V_{gb}\ne
0)/LDOS(V_{gb}=0)$. $LDOS(V_{gb}=0)$  contains the 
inhomogeneous charge distribution and $LDOS(V_{gb}\ne 0)$ vanishes 
around the Fermi energy due to the superconducting 
gap and the single particle levels in the grains. 
This LDOS ratio yields well-defined peaks 
and converges to the unity at large bias,
in close agreement to similar LDOS ratios calculated from STM 
measurements\cite{Pasupathy}.

\begin{figure}[ht]
\begin{center}
%\begin{minipage}[b]{.1\textwidth}
%\begin{center}
%     \centerline{\includegraphics[width=6.0cm,angle=-90]{GnuP.EjxT4.ps}}
     \centerline{\includegraphics[width=8.0cm]{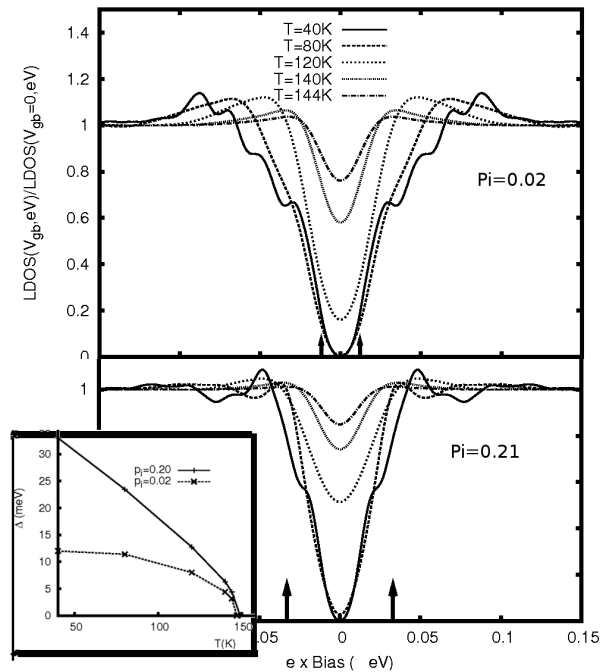}}
%    \vspace{0.45cm}
%    \centerline{\includegraphics[width=5.0cm,angle=-90]{LDOSV7T09.n21Ts.461.ps}}
%   \end{center}
%  \end{minipage}
\caption{The $p=0.11$ LDOS. 
Top panel, $\Delta_{PG}(T)$ at a hole-poor puddle
($p_i=0.021$). %pi=461
$\Delta_{PG}(T=40K)\approx 80$meV. The superconducting 
gaps $\Delta_d(T=40K)=12$meV are marked by arrows.  
Below, the set of LDOS curves at a hole-rich
grain, ($p(i)\approx 0.20$) with  $\Delta_{PG}(T=40K)\approx 50$meV
and with $\Delta_d(T=0)=33meV$. %pi=471
In the inset $\Delta_d(i,T) \times T$ for each case. Both
$\Delta_{PG}(i)$ and $\Delta_{d}(i)$ vanish near $T=147$K.} 
\label{LDOSn11} 
\end{center}
\end{figure}

We calculated the LDOS at many values of $p$ and $T$ to compare
with the STM\cite{Pan,McElroy,Gomes,Pasupathy,Kato,Pushp} 
and ARPES\cite{Shen,Mad,Campuzano,Chatterjee} 
data, but here we present detailed 
results only for an underdoped ($p=0.11$)
and an overdoped ($p=0.21$) sample. Thus, in Fig.(\ref{LDOSn11}) 
we  show the results at two representative
places (shown as white dots in the middle of Fig.(1)) 
of the $p=0.11$ sample at various temperatures. 
In this case, the low temperature LDOS
display very few well defined (coherent) peaks but
the gaps are similar to the STM data 
of McElroy et al\cite{McElroy}. The low temperature
superconducting gap $\Delta_d$ produces small anomalies
that are marked by arrows, as reported by some 
STM data\cite{Kato,Pushp,Kato2}. This behavior was 
already noticed by our
previous work\cite{Mello09}. Then the gaps derived from the 
LDOS peaks are identified with the local pseudogap $\Delta_{PG}(i,p,T)$ 
because they are larger than the local
$\Delta_{d}(i,p,T)$ and because they vanish
at a higher temperature $T^*(i)$ than $T_c(i)$.

Let us investigate in detail the 
relationship between $\Delta_{PG}(i,p,T)$ and $\Delta_{d}(i,p,T)$.
For the  $p=0.11$ compound with $T_c(0.11)\approx 65$K
at a typical hole-poor place with $p_i=0.02$, we obtain a 
low temperature $\Delta_{PG}\approx 80$meV  and  $\Delta_d=12$meV. 
In a representative hole-rich location with $p(i)\approx 0.23$  there is an 
interchange of intensity: $\Delta_{PG}\approx 50$meV and the 
superconducting gap $\Delta_d(T=0)=33$meV. 
We observe that this inversion in the values of $\Delta_{PG}$ 
and $\Delta_{d}$ is common to
all compounds. For this $p=0.11$ sample, at all sites, 
$\Delta_{PG}(i)$ and $\Delta_{d}(i)$ vanish with increasing 
temperature almost together near $T=145$K. At the  $p_i=0.02$
hole poor grain, $\Delta_{PG}$ and $\Delta_d$ vanish near 
$T^*(i)=T_c(i)=147$K, and at the $p_i=0.21$ hole rich grain, they 
close at $T^*(i)=148$ and $T_c(i)=146$K. We see that, 
for underdoped samples, $T^*(i,p) \approx T_c(i,p)$ and they
remain much above the resistivity transition 
$T_c(0.11)\approx 65$K, i.e., {\it both gaps are present
in the pseudogap phase up to $T^*$.}

\begin{figure}[ht]
\begin{center}
%\begin{minipage}[b]{.1\textwidth}
%\begin{center}
%     \centerline{\includegraphics[width=6.0cm,angle=-90]{GnuP.EjxT4.ps}}
     \centerline{\includegraphics[width=8.0cm]{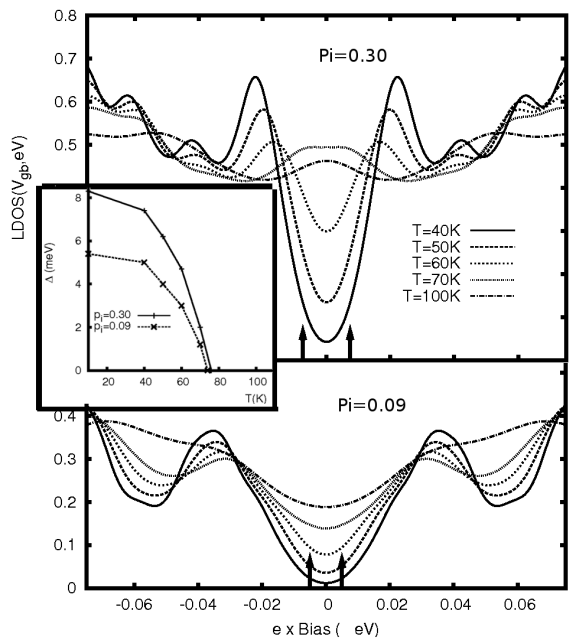}}
%    \vspace{0.45cm}
%    \centerline{\includegraphics[width=5.0cm,angle=-90]{LDOSV7T09.n21Ts.461.ps}}
%   \end{center}
%  \end{minipage}
\caption{LDOS for overdoped $p=0.21$.
At the top, $\Delta_{PG}(T)$ at a hole-rich grain ($p_i\approx 0.30$) 
yielding $\Delta_{PG}(T=40K)\approx 24$meV 
%(i=413)
and $\Delta_d(T=40K)=7meV$ marked by the arrows.  Below,
$\Delta_{PG}(T)$ at an "insulator" grain
%(i=461)
($p(i)\approx 0.09$), $\Delta_{PG}(40K)\approx 37$meV and
$\Delta_d(T=40K)=4.5 meV$. In the inset, we show also that $\Delta_d(T)\times T$ 
at the two locations. It is important
to notice that $\Delta_{PG}(i)$ remains even above $T=100$K$>>T_c(i)$ for
some hole-poor grains $i$. } 
\label{LDOSn21} 
\end{center} 
\end{figure}

The difference between $T^*(i,p)$ and $T_c(i,p)$ 
increases continuously with increasing doping. Calculations with
$p=0.16$ show that in some grains $T^*(i) \approx T_c(i)+15$K
while in some other locations $T^*(i)\approx T_c(i)$.
By increasing the doping $p$, this difference increases 
as shown in Fig.({\ref{LDOSn21}) for $p=0.21$.
For this compound with $T_c(0.21)\approx 72$K, 
most of the LDOS present well 
defined coherent peaks as observed by McElroy et al\cite{McElroy}. On
the top panel, we show the LDOS at a hole-rich site
$p_i\approx 0.30$; we see that $\Delta_{PG}(T=40K)\approx 24$meV and
$\Delta_d(T=40K)=7.1$meV (marked by arrows in the plots) 
and they both vanish at 
$T^*(i)\approx T_c(i) \approx 72K$. At a hole-poor site 
of $p_i=0.9$, $\Delta_{PG}(T=40K)\approx 37$meV and
$\Delta_d(T=40K)=4.5$meV. Following the temperature 
evolution we see that $T_c(i)\approx 72K$ that
is also the critical temperature $T_c(0.21)$, and
$\Delta_{PG}$ vanishes at much larger value $T^*(i)\approx 105K$.
Thus, {\it the pseudogap phase of overdoped samples is composed mainly
by $\Delta_{PG}$}.

The above results lead us to many conclusions
concerning the HTSC measured properties: 
{\it i}) The charge inhomogeneities and the charge segregation potential
occur due to the formation of (almost zero doping) AF regions. 
Consequently, its effect is more intense at low doping compounds, 
leading to large superconducting amplitudes $\Delta_d$ and 
$\Delta_{PG}$ in underdoped samples. On the other hand,
at far overdoped region, the large Coulomb
repulsion in the hole-rich grains destroys the EPS 
transition and the superconductivity.
{\it ii}) As in granular superconductors, the resistivity 
transition occurs by Josephson coupling  among the intragrain 
superconducting regions.
{\it iii}) In this scenario, the pseudogap is due to
the weakly localized energy levels in the
two dimensional puddles and it is not directly related to the 
intragrain superconductivity, although both phenomena are 
originated by the segregation potential 
$V_{gb}$. This different process was demonstrated by the
distinct behavior of the two signal with 
applied magnetic fields and also with the temperature 
in tunneling experiments\cite{Krasnov}.
{\it iv}) The observed difference in the
LDOS form, called "coherent" and "zero temperature 
pseudogap"\cite{McElroy}, is due to the LDOS at hole-rich and 
hole-poor locations respectively. This distinction is clearly 
seen in the overdoped 
calculations (Fig.(\ref{LDOSn21})) where most LDOS presents
coherent peaks in opposition to the underdoped case (as in 
Fig.(\ref{LDOSn11})) where the rounded and 
ill-defined peaks are more abundant. 
{\it v}) The hole-poor regions have basically one electron per
unit cell and the large Coulomb repulsion generates a large
asymmetry between electron extraction and injection as measured
by the STM experiments\cite{Pan,McElroy,Gomes,Pasupathy,Kato,Pushp,Kato2}.
The LDOS at hole-rich regions, like a homogeneous system, are expected to
be more symmetric. {\it vi}) The observed anomaly or kink measured in the LDOS at very low 
bias\cite{Kato,Pushp,Kato2} was demonstrated to be due to the low temperature
superconducting gap $\Delta_{d}$. Fig.(\ref{LDOSn11}) 
shows clearly a small kink marked by the arrows. In general, it is
more visible in underdoped samples because the $\Delta_{PG}$
is much larger than  $\Delta_{d}$. 
{\it vii}) For underdoped samples, as in Fig.(\ref{LDOSn11}), 
$\Delta_d$ and $\Delta_{PG}$  remain finite much above and
vary very little around $T_c(p)\approx 65$K. 
The presence of the superconducting gap and 
its quasiparticle dispersion above
$T_c(p)$ was measured in weakly underdoped Bi2212 by
ARPES experiments\cite{Campuzano} that also showed that
the gap ($\Delta_{PG}$) almost does not change around $T_c(p)$. 
{\it viii}) The ARPES experiment of Lee et al\cite{Shen} has
measured increasing gapless Fermi arcs along the nodal ($\pi,\pi$) 
region with increasing doping above $T_c(p)$. As we have shown, 
the d-wave $\Delta_d$ remains above $T_c(p)$ in the underdoped 
regions but tends to vanish close to $T_c(p)$ for overdoped compounds. 
Assuming that the weakly bound states ($\Delta_{PG}$) are
due to the random phase boundary potential $V_{gb}$, 
and that they occur mainly along the  $Cu-O$ (antinode) direction,
together with the d-wave behavior of $\Delta_d$ is agreement with
the increase of the gapless Fermi arcs with $p$
above $T_c(p)$\cite{Shen}. {\it ix}) The calculations show that, in general, for any compound, 
the pseudogap $\Delta_{PG}$ is smaller but with
well defined peaks at the hole-rich regions than in
the hole-poor ones. Consequently, the measured larger
$\Delta_{PG}(i)$ gap values located at the low density grains with
local (AF) insulator behavior have lower local conductivity 
($dI/dV$) than the smaller gaps (and higher densities)  as 
verified by Pasupathy et al\cite{Pasupathy}. {\it x}) The STM measured relation
$2\Delta/K_BT^*(i) \approx 8.0$\cite{Gomes} is reproduced
closely by the optimal and the $p=0.21$ sample, considering
$\Delta=\Delta_{PG}(i)$ and its associated $T^*(i)$.

In summary, we have proposed an EPS transition
to describe the inhomogeneous charge distribution of HTSC 
generated by the lower free energy of the (undoped) AF domains. 
This approach yields potential wells with shallow bound states
that reduce the kinetic energy and favor the superconducting
pairing. The calculations obtain the inverted bell shape
critical line $T_c(p)\times p$, 
distinguish clearly the LDOS pseudogap $\Delta_{PG}$   
from the intragrain superconducting gap energy 
$\Delta_{d}$ and
provide simple physical interpretations to 
many different STM and ARPES results.

We gratefully acknowledge partial financial aid from Brazilian
agency CNPq. 
%Discussions with J.C. Campuzano and J.M. Tranquada
%are gratefully acknowledged.
%


\begin{references}
%\begin{thebibliography}
%
\bibitem{Tranquada} J.M.Tranquada, et al 
%B.J. Sternlieb, J.D. Axe, Y. Nakamura, and S. Uchida, 
Nature (London),{\bf 375}, 561 (1995).
%%
%\bibitem{Bozin} E. S. Bozin, G. H. Kwei, H. Takagi, and S. J. L. Billinge,
%Phys. Rev. Lett.  {\bf 84}, 5856 (2000).
%
%\bibitem{Uemura} Y.J. Uemura, Sol. Stat. Comm. {\bf 126}, 23 (2003).
%
%\bibitem{Singer} P. M. Singer, A. W. Hunt, and T. Imai,
%Phys. Rev. Lett. {\bf 88}, 47602 (2002).
%
%\bibitem{Curro} H.-J. Grafe, N. J. Curro, M. H\"ucker, and B. B\"uchner,
%Phys. Rev. Lett., {\bf 96}, 017002 (2006).
%
%
\bibitem{Pan}  S. H. Pan et al,
Nature (London), 413, 282-285 (2001).
%%
%\bibitem{Lang} K.M. Lang et al,
%% V. Madhavan, J.E. Hoffman, E.W. Hudson, H. Eisaki, S. Uchida
%and J.C. Davis,
%Nature, {\bf 415}, 412 (2002).
%
\bibitem{McElroy} K. McElroy, et al
%D.-H. Lee, J. E. Hoffman, K. M Lang,
%E. W. Hudson, H. Eisaki, S. Uchida, J. Lee, J.C. Davis,
cond-mat/0404005 an Phys. Rev. Lett. {\bf 94}, 197005 (2005).
%
\bibitem{Gomes} Kenjiro K. Gomes et al,
%Abhay N. Pasupathy, Aakash Pushp, Shimpei Ono,
%Yoichi Ando, and Ali Yazdani,
Nature {\bf 447}, 569 (2007).
%
\bibitem{Pasupathy} Abhay N. Pasupathy et al,
%Kenjiro K. Gomes, Colin V.Parker, Jinsheng Wen, Zhijun Xu,
%Genda Gu, Shimpei Ono, Yoichi Ando, Ali Yazdani,
Science {\bf 320}, 196 (2008).
%
%\bibitem{Alldredge} J. W. Alldredge et al,
%Nature Phys. {\bf 4}, 319 (2008).
%
\bibitem{Kato} Takuya Kato et al, J. Phys. Soc. Jpn., {\bf 77}, 054710 (2008).
%
\bibitem{Pushp} Aakash Pushp et al, Science {\bf 324} 1689 (2009).
%%
\bibitem{Kato2} T. Kato et al, J. Supercond. Nov. Magn. {\bf 23}, 771 (2010).
%\bibitem{Gorgov} L. P. Gor`kov and A. V. Sokol, JETP Lett. {\bf 46}, 420 (1987).
%
%\bibitem{Yukalov} V. I. Yukalov and E. P. Yukalova,
%Phys. Rev. {\bf B70}, 224516 (2004).
%
%
\bibitem{LeTacon} M. Le Tacon, et al, 
% A. Sacuto, A. Georges, G. Kotliar, Y. Gallais, D. Colson and  A. Forget, 
Nature Phys. {\bf 2}, 537 (2006).
%
\bibitem{Shen} W. S. Lee, et al, Nature {\bf 450}, 81 (2007)
%
\bibitem{Mad} J. H. Ma, et al, Phys. Rev. Lett. {\bf 101}, 207002 (2008).
%
\bibitem{Campuzano} A.Kanigel et al,
% U. Catterjee, M. Randeria, G. Koren,
%K. Kadowaki, and J. C. Campuzano,
Phys. Rev. Lett. {\bf 101}, 137002 (2008).
%
\bibitem{Chatterjee} U. Chatterjee, et al  Nature Phys. {\bf 6}, 99 (2010).
%
%\bibitem{Zaanen} J. Zaanen and O. Gunnarsson, Phys. Rev. {\bf B40}, 7391 (1989).
%
%\bibitem{EK}V. J. Emery and S. A. Kivelson, Physica {\bf C209}, 597 (1993).
%%
\bibitem{Mello09} E.V.L. de Mello, et al
J. Phys.: Condens. Matter {\bf 21}, 235701 (2009).
%
\bibitem{Tranquada2} S. Wakimoto at al  Phys. Rev. Lett. {\bf 98}, 247003 (2007)
%\bibitem{Mello03} E.V.L. de Mello et al,
%E.S. Caixeiro, and J.L. Gonz\'alez,
%Phys. Rev. {\bf B67}, 024502 (2003).
%
\bibitem{CH} J.W. Cahn and J.E. Hilliard, J. Chem. Phys, {\bf 28}, 258
(1958).
%
\bibitem{Otton} E.V.L de Mello et al,
Physica {\bf A 347}, 429 (2005).
%
%
\bibitem{Mello04} E.V.L. de Mello et al,
 Phys. Rev. {\bf B70}, 224517 (2004).
%
\bibitem{Landau} L.D. Landau, and E.M. Lifshitz, ``Quantum Mechanics'',
Pergamon Press, New York, 1977.
%
\bibitem{JR} E. V. L. de Mello, et al, Phys. Rev. {\bf
B58}, 9098 (1998).
%
%
%\bibitem{Bray} A.J. Bray, Adv. Phys. {\bf 43}, 347 (1994).
%
%\bibitem{DDias07} E. V. L. de Mello, and D. N. Dias, J. Phys.
%C.M. {\bf 19}, 086218 (2007).
%
\bibitem{DDias08} D. N. Dias et al, Phys. {\bf C468}, 480 (2008).
%
\bibitem{Mello08} E.V.L. de Mello et al,
% R. B. Kasal, C.A.C. Passos, e Otton S.T. Filho,
Physica {\bf B404},  3119 (2009).
%
%\bibitem{Caixa07} E. S. Caixeiro, E.V.L. de Mello, and A. Troper,
%Physica {\bf C459}, 37 (2007).
%
\bibitem{Merchant} L. Merchant et al, Phys. Rev. {\bf B63}, 134508 (2001).

%\bibitem{JL07} D. N. Dias et al, Phys. Rev. {\bf B76}, 90737 (2007).
%
%\bibitem{Stanley} H. E. Stanley, "Introductin to Phase Transitions
%and Critical Phenomena" (Oxford U. Press, N.Y., 1971).
%
%
%\bibitem{Loram} J.W. Loram et al,
%J. Luo, J. R. Cooper, W.Y. Liang,
%and J.L. Tallon ,
%J. Phys. Chem. Sol. {\bf 62}, 59 (2001).
%
%\bibitem{Hardy}J. E. Sonier et al , Phs. Rev. Lett. {\bf 101}, 117001
%(2008).
%
%\bibitem{TS} T. Timusk and B. Statt, Rep. Prog. Phys., {\bf 62}, 61 (1999).
%
%
%\bibitem{Tallon} J.L. Tallon and J.W. Loram, Physica C {\bf 349}, 53 (2001).
%
%\bibitem{Lee} Patrick A. Lee, Naoto Nagaosa, and Xiao-Gang Wen,
%Rev. Mod. Phys. {\bf 78}, 17 (2006).
%
\bibitem{AB} V. Ambeogakar, and A. Baratoff, Phys. Rev. Lett. {\bf 10},
486 (1963).
%
%
\bibitem{Takagi} H. Takagi et al, Phys. Rev. Lett. {\bf 69}, 2975 (1992).
%
\bibitem{Gygi} Fran\c{c}ois Gygi, and Michael Schl\"uter, Phys. Rev. {\bf B43}, 7609 (1991).
%
%\bibitem{Tranquada2} S. Wakimoto et al,
%K. Yamada, J. M. Tranquada, C. D. Frost,
%R. J. Birgeneau, and H. Zhang,
%Phys. Rev. Lett. {\bf 98}, 247003 (2003).
%
%\bibitem{Suzuki} Minoru Suzuki, and Takao Watanbe, Phys. Rev. Lett.,
%{\bf 85}, 4787 (2000).
%
%\bibitem{Ong} Yayu Wang, Lu Li, and N. P. Ong, Phys. Rev. B {\bf 73},
%024510 (2006).
%
\bibitem{Krasnov} V. M. Krasnov et al, Phys. Rev. Lett. {\bf 86}, 2657 (2001).
%\bibitem{Tallon2} J. L. Tallon et al, Phys. Stat. Sol {\bf 215}, 531 (1999).
%
%\bibitem{Kohsaka} Y. Kohsaka et al, Science {\bf 315}, 1380 (2007).
%
%\bibitem{Dynes} R. C. Dynes, V. Narayanamurti, and J. P. Garno,
%Phys. Rev. Lett. {\bf 41}, 1509 (1978).
%
%\bibitem{Dagotto} E. Dagotto et al, Sol. Stat. Comm. {\bf 126}, 9 (2003).
%
\end{references}
\end{document}